\documentstyle[pre,aps,epsfig,amssymb]{revtex}

\begin{document}
\draft

\title{Hard-Sphere Fluids in Contact with Curved Substrates}
\author{P. Bryk,$^{1,2,3}$ R. Roth,$^{2,3}$, K.R. Mecke,$^{2,3}$,
and S. Dietrich,$^{2,3}$}

\address{
$^1$ Department for the Modeling of Physico-Chemical Processes, 
Maria Curie-Sk{\l}odowska University, 20-031 Lublin, Poland\\ 
$^2$ Max-Planck-Institut f{\"u}r Metallforschung,
Heisenbergstrasse 3, D-70569 Stuttgart, Germany\\
$^3$ Institut f{\"u}r Theoretische und Angewandte Physik,
Universit{\"a}t Stuttgart, Pfaffenwaldring 57, D-70569 Stuttgart, Germany}

\maketitle
\begin{abstract}
The properties of a hard-sphere fluid in contact with hard spherical and 
cylindrical walls are studied. Rosenfeld's density functional theory (DFT) 
is applied to determine the density profile and surface tension $\gamma$ for
wide ranges of radii of the curved walls and densities of the hard-sphere 
fluid. Particular attention is paid to investigate the curvature dependence 
and the possible existence of a contribution to $\gamma$ that is
proportional to the logarithm of the radius of curvature. Moreover, by treating
the curved wall as a second component at infinite dilution we provide an
analytical expression for the surface tension of a hard-sphere fluid close to
arbitrary hard convex walls. The agreement between the analytical expression 
and DFT is good. Our results show no signs for the existence of a logarithmic 
term in the curvature dependence of $\gamma$. 
\end{abstract}

\section{Introduction}
The properties of inhomogeneous fluids with spherical or cylindrical geometries
have been investigated by many authors in recent years 
\cite{Poniewierski97,Samborski93,Stecki90,Bieker98,Bauer00,Degreve94,Dhenderson95,Attard01,Giessen98} 
because curvature plays an important role in many physical situations, including
the following examples. In the
context of nucleation phenomena \cite{Oxtoby92,Wolde98,Barret99} one considers
the spontaneous formation of spherical droplets in a super-saturated vapor. The
variation of the local curvature  of confinements leads to a more complicated 
form of the depletion forces in colloidal suspensions 
\cite{Roth99,Roth00,Roth01}. Wetting or critical adsorption of a fluids on 
curved substrates, such as spherical or cylindrical colloidal particles
may lead to the formation of liquid bridges if two colloids come close to each
other, and subsequently to flocculation \cite{Bieker98,Dhenderson95,Beysenes94,Archer02,Schlesener03,Hanke99}.
In studies of the hydrophobic effect one is concerned with the cost in free 
energy to immerse a hydrophobic particle into a liquid \cite{Lum99,Katsov01}.
Despite these efforts, the properties of curved fluid interfaces are not 
understood as fully as the planar one. Whether one deals with a drop of 
liquid surrounded by its vapor or with a big spherical particle 
immersed into a solvent of small particles, the statistical-mechanical description of such systems generates 
additional difficulties \cite{Schofield82,Henderson82} which call for further 
studies \cite{Henderson83,Henderson84,Henderson86}.

Within this context, here we consider the case of a hard-sphere fluid in the grand 
canonical ensemble, characterized by the radius $R$ of the spheres and the
bulk number density $\rho$ or equivalently the bulk packing fraction 
$\eta=\left (4 \pi/3\right ) R^3\rho$, in contact with a hard spherical or 
cylindrical wall. These curved walls can either be viewed as an external 
potential exerted on the hard-sphere fluid, 
\begin{equation}\label{eqn:1}
V^{ext}_{i}(r)=\left\{
\begin{array}{ll}
0, & r>R_{i} \\
\infty, &  r<R_{i} \;,
\end{array}
\right.
\end{equation}
where the subscript $i=s$, $c$ denotes the spherical and cylindrical symmetry,
respectively, or as the surfaces of a second ({\it b}ig) species of radius 
$R_i^{(b)}$. The radii $R_i$ entering the external potential in 
Eq.~(\ref{eqn:1}) are related to the actual radii $R_i^{(b)}$ via 
$R_i=R_i^{(b)}+R$ (see Fig.~\ref{fig:geometry}).

The grand canonical potential of the system $\Omega_i$ can be divided into a 
bulk part $\Omega^{bulk}_i=-p V_{i}$, where $p$ is the pressure of the 
hard-sphere fluid and $V_{i}$ is the volume accessible to the system, and an 
excess (surface) part $\Omega^{ex}_i$: 
$\Omega_{i}=\Omega^{bulk}_i+\Omega^{ex}_i$. The surface tension $\gamma_{i}$ 
is readily {\it defined} as the excess grand potential per unit area,
\begin{equation}\label{eqn:2}
\gamma_i=\frac{\Omega^{ex}_i}{A_i^{(b)}}=\frac{\Omega_i-\Omega^{bulk}_i}{A_i^{(b)}}\;,
\end{equation}
where $A_{i}^{(b)}$ is the surface area of the curved wall. 
While $\Omega_i$ is uniquely defined this division into bulk and surface excess
parts is not unique. Depending on the choice of the dividing surface the 
surface tension can even change sign. As indicated by the superscript $(b)$ 
in Eq.~(\ref{eqn:2}) we use the actual wall corresponding to the radius
$R_i^{(b)}$ as the dividing interface.

Within density functional theory (DFT), the grand potential $\Omega_i$ of the 
system can be expressed as
\begin{equation}\label{eqn:5}
\Omega_{i}[\rho]={\cal F}[\rho]+\int d^3 r\rho({\bf r})
(V^{ext}_{i}({\bf r})-\mu)\;,
\end{equation}
where ${\cal F}[\rho]={\cal F}_{ex}[\rho]+{\cal F}_{id}[\rho]$ is the 
intrinsic free energy functional that can be split into an ideal gas
contribution ${\cal F}_{id}[\rho]$ and an excess (over the ideal gas) term
${\cal F}_{ex}[\rho]$. The only explicit dependence of $\Omega_i[\rho]$ on the
external potential originates from the second term in Eq.~(\ref{eqn:5}). 
The equilibrium density profile $\rho({\bf r})$ satisfies the Euler-Lagrange equation
\begin{equation}\label{eqn:4}
\frac{\delta \Omega_{i}[\rho]}{\delta \rho({\bf r})}\; = 0 \;.
\end{equation}
For the equilibrium density profile $\rho({\bf r})$ the density functional
reduces to the grand potential of the system, i.e., 
$\Omega_i=\Omega_i[\rho({\bf r})]$, and we can rewrite Eq.~(\ref{eqn:2}) as
\begin{equation}\label{eqn:3}
\gamma_{i}=\frac{\Omega_{i}[\rho({\bf r})]+p V_{i}}{A_{i}^{(b)}}\;.
\end{equation}
From Eqs.~(\ref{eqn:1}), (\ref{eqn:5}), and (\ref{eqn:4}) it follows that the 
change in the grand potential due to an infinitesimal change of the radius 
$R_{i}^{(b)}$ of the curved wall, at constant chemical potential $\mu$ and
temperature $T$, is given by
\begin{equation}\label{eqn:6}
\left(\frac{\partial \Omega_i}{\partial R_i^{(b)}}\right)_{\mu,T}=\int d^3 r
\frac{\delta \Omega_i[\rho]}{\delta \rho({\bf r})}
\frac{\partial \rho({\bf r})}{\partial R_i^{(b)}}+\int d^3 r \rho({\bf r})
\frac{\partial V_i^{ext}({\bf r})}{\partial R_i^{(b)}}\;.
\end{equation}
The first term on the right-hand side of Eq.~(\ref{eqn:6}) vanishes by virtue 
of Eq.~(\ref{eqn:4}) while the second term, for the external potential given in
Eq.~(\ref{eqn:1}), is equal to $\beta^{-1} A_i~\rho(R_i^+)$, where 
$\rho(R_i^+)$ denotes the contact density of the hard-sphere fluid at the 
curved wall. This leads to the sum rules \cite{Henderson83}
\begin{equation}\label{eqn:7}
\beta \left(\frac{\partial \Omega_i}{\partial R_i^{(b)}} \right)_{\mu,T}=
\left\{
\begin{array}{ll}
4\pi R_s^2\rho(R_{s}^+),& \mbox{for a spherical wall}\\
2\pi R_c L\rho(R_c^+),&\mbox{for a cylindrical wall.}
\end{array}
\right.
\end{equation}
In the case of a cylindrical wall we consider the thermodynamic limit
$\displaystyle{\lim_{L\to\infty} (\Omega_c/L)}$ so that effects due to a 
finite length $L$ of the cylinder drop out. We note that the sum rules in 
Eq.~(\ref{eqn:7}) are valid for all one-component fluids in contact with hard 
spherical or cylindrical walls and are satisfied by all density functionals
within weighted-density approximation (WDA) \cite{Samborski93}.

Using the actual radius  $R_i^{(b)}$ also as the radius of the dividing 
interface (see Fig.~\ref{fig:geometry}), the grand potential 
$\Omega_i=-p V_i + \gamma_i A_i^{(b)}$ of the system is separated into a bulk 
and surface term with the accessible volume
\begin{equation}\label{eqn:8}
V_i=V_{i}(\overline{R})-V_{i}(R_i^{(b)}) \;,
\end{equation}
where $V_{s}(r)=\frac{4\pi}{3}r^{3}$, $V_{c}(r)=\pi r^{2}L$, and 
$\overline{R}$ is a macroscopicly large radius considered 
in the thermodynamical limit, and the surface areas
\begin{equation}\label{eqn:9}
A_i^{(b)}=\left\{
\begin{array}{ll}
4\pi (R_{s}^{(b)})^{2}, & \mbox{for a spherical wall} \\
2\pi R_{c}^{(b)}L, &  \mbox{for a cylindrical wall}\;.
\end{array}
\right.
\end{equation}
The sum rule in Eq.~(\ref{eqn:7}) can then be expressed as
\begin{equation}\label{eqn:10}
\rho(R_{s}^+) \left(\frac{R_s}{R_s^{(b)}}\right)^2 =\beta p+
\frac{2 \beta \gamma_{s}(R_{s}^{(b)})}{R_{s}^{(b)}}+\beta
\left(\frac{\partial\gamma_{s}(R_{s}^{(b)})}{\partial R_{s}^{(b)}}\right)_{\mu,T}
\end{equation}
for a spherical wall and
\begin{equation}\label{eqn:11}
\rho(R_{c}^+) \frac{R_c}{R_c^{(b)}}=\beta p+\frac{\beta\gamma_{c}(R_{c}^{(b)})}{R_{c}^{(b)}}+
\beta\left(\frac{\partial\gamma_{c}(R_{c}^{(b)})}{\partial R_{c}^{(b)}}\right)_{\mu,T}
\end{equation}
for a cylindrical wall. Equations~(\ref{eqn:10}) and (\ref{eqn:11}) indicate 
that different versions of DFTs might lead to different contact values near 
{\it curved} walls even if the underlying equation of state is the same, which
leads to identical contact values at a hard {\it planar} wall. As pointed out 
by J.R. Henderson \cite{Henderson86}, these sum rules have interesting 
implications. If  $\rho(R_i^+)$ is analytic as a function of 
$R_i^{-1}=(R_{i}^{(b)}+R)^{-1}$ one has
\begin{equation}\label{eqn:12}
\rho(R_i^+)=c_0+\frac{c_i^{(1)}}{R_i^{(b)}+R}+\frac{c_i^{(2)}}{(R_i^{(b)}+R)^2}+
\frac{c_i^{(3)}}{(R_i^{(b)}+R)^3}+
\cdots
\end{equation}
with $c_0=\beta p$ for both the spherical and the cylindrical wall and 
expansion coefficients $c_i^{(j)}$ for $j=1,2,\ldots$. Note that the expansion
in Eq.~(\ref{eqn:12}) is assumed to converge for arbitrary non-zero values of 
$R_{i}$. This assumption will be corroborated by our numerical results in 
Sec.~\ref{sec:DFT}. With Eq.~(\ref{eqn:12}) the solution of the differential
equation (\ref{eqn:10})  can be written as
\begin{equation}\label{eqn:13}
\beta \gamma_{s}(R_s^{(b)})=\frac{C_s^{(1)}}{2}\left [1+\frac{2 C_s^{(2)}}{C_s^{(1)}}
\frac{1}{R_s^{(b)}}+\frac{2 c_s^{(3)}}{C_s^{(1)}}\frac{\ln (R_s^{(b)}/R)}{(R_s^{(b)})^2}
+\frac{2 D_s}{C_s^{(1)}}\frac{1}{(R_s^{(b)})^2}+{\cal O}\left( \frac{1}{(R_s^{(b)})^3}\right)
\right ]\;\;,
\end{equation}
where $D_{s}$ is an integration constant, $C_{s}^{(1)}=c_{s}^{(1)}+2\beta pR$
and $C_{s}^{(2)}=c_{s}^{(2)}+c_{s}^{(1)}R+\beta pR^{2}$. Likewise, with 
Eq.~(\ref{eqn:12}) the solution of the differential equation (\ref{eqn:11}) can be
written as
\begin{equation}\label{eqn:14}
\beta \gamma_{c}(R_c^{(b)})=C_c^{(1)}\left[1+\frac{c_c^{(2)}}{C_c^{(1)}}
\frac{\ln (R_c^{(b)}/R)}{R_c^{(b)}}+\frac{D_c}{C_c^{(1)}}\frac{1}{R_c^{(b)}}+
{\cal O}\left(\frac{1}{(R_c^{(b)})^2}\right )\right ]\, ,
\end{equation}
where $D_{c}$ is an integration constant and 
$C_{c}^{(1)}=c_{c}^{(1)}+\beta pR$. In the case of the spherical wall 
$\gamma_{s}(R_{s}^{(b)})$ has the form
\begin{equation}
\beta\gamma_{s}(R_{s}^{(b)})=\beta\gamma_{s}(\infty)\left [ 1-\frac{2\delta_{T}}{R_{s}^{(b)}}
+{\cal O}\left(\frac{\ln(R_{s}^{(b)}/R)}{(R_{s}^{(b)})^{2}}\right)
\right ]
\end{equation}
where $\gamma_{s}(\infty)=\beta^{-1}\frac{c_{s}^{(1)}}{2}+pR$ is the surface 
tension of the planar wall and 
\begin{equation}
\delta_{T}=-\frac{c_{s}^{(2)}+c_{s}^{(1)}R+\beta pR^{2}}{c_{s}^{(1)}+2\beta pR}
\end{equation}
is the so-called Tolman length \cite{Tolman49}, which plays an  important role
in nucleation theory \cite{Bieker98,Oxtoby92,Wolde98,Barret99}. 
We note that the value of the Tolman length depends on the choice of
the dividing surface; here the dividing surface has the radius $R_{s}^{(b)}$.
If these logarithmic 
terms in the expansion of the surface tensions do not vanish, the concept of 
the Tolman length apparently breaks down for cylindrically curved and 
generally shaped walls \cite{Henderson84,Henderson86}.

The analyticity of $\gamma(R)$ in terms of $\frac{1}{R}$ is of considerable
importance because it provides the basis for the so-called Helfrich expansion 
of the surface free energy of arbitrarily curved surfaces in powers of the 
principal curvatures \cite{Helfrich73,David89}. Although this approach appears
to be very useful for describing membranes \cite{Lipowsky95} it has been shown
recently that the presence of long-ranged dispersion forces destroys the 
aforementioned analyticity for fluid interfaces \cite{Bieker98,Mecke99,Mecke00}
and prevents the use of a Helfrich-type interface Hamiltonian for such systems.
The possible occurrence of  logarithmic terms in Eqs. (\ref{eqn:13}) 
and (\ref{eqn:14}) following from the simple expansion (\ref{eqn:12})
raises the issue whether the absence of analyticity is not only due 
the presence of long-ranged forces but is an intrinsically geometrical 
effect which shows up 
even for surface free energies of hard-sphere fluids. The present 
study aims at clarifying this conceptual point.

To this end we investigate the curvature dependence of the surface tension as 
well as relationships between thermodynamic and local properties of the 
hard-sphere fluids following two different routes. First, in 
Sec.~\ref{sec:bulk} the surface tension is determined from a bulk theory in 
which the curved wall is considered as the surface of an additional component 
at infinite dilution. Second, a numerical approach based on the minimization 
of the Rosenfeld fundamental measure theory (FMT) free energy functional 
\cite{Rosenfeld89,Rosenfeld93} is presented in Sec.~\ref{sec:DFT}. In this 
approach, the hard-sphere fluid is considered to be exposed to the external 
potential of Eq.~(\ref{eqn:1}). 

\section{Bulk Theory} \label{sec:bulk}
We start by considering the bulk of a one component hard-sphere fluid in a 
volume $V_{tot}$ with $V_{tot}\to {\mathbb R}^{3}$ in the thermodynamic limit. 
The grand potential of this system is given by $\Omega_0=-p V_{tot}$, where $p$
is the pressure of the hard-sphere fluid. A single particle of a second 
component $b$ is inserted into this system. In view of the following
discussion the only restriction for this 
particle is that it must be hard and convex. The grand potential of the new 
system is 
\begin{equation} \label{eqn:b1}
\Omega=\Omega_0+\Delta\Omega \;, 
\end{equation}
where $\Delta\Omega$ measures the change of the grand potential due to the 
insertion of a single particle. $\Delta\Omega$ equals the one-particle direct 
correlation function \cite{Roth00,Roth01,Dijkstra99} which in the bulk limit 
considered here is simply the excess chemical potential of species $b$ in the 
dilute limit $\rho_{b}\to 0$, i.e.,
\begin{equation}\label{eqn:15}
\beta\Delta\Omega=-c_{b}^{(1)}\;.
\end{equation}
Although Eqs.~(\ref{eqn:b1}) and (\ref{eqn:15}) are formally exact it is 
important to notice that this approach within a bulk theory is expected to 
only work reliably if the considered fluid is in a single phase. In the case 
of phase separation, which could lead to wetting or drying of the big sphere,
this approach most likely would fail \cite{Henderson02}.

We evaluate Eq.~(\ref{eqn:15}) within the framework of Rosenfeld's Fundamental 
Measure Theory (FMT). Within this approach \cite{Rosenfeld89,Rosenfeld93} the 
excess free energy functional (over the ideal gas) is given by
\begin{equation}\label{eqn:20}
\beta {\cal F}_{ex}=\int d^{3}r \;\Phi(\{n_{\alpha}\}) \;,
\end{equation}
where $n_{\alpha}$ denote weighted densities 
\begin{equation}\label{eqn:21}
n_{\alpha}({\bf r})=\sum_{j=1}^{N}\int d^{3}r' \rho_j({\bf r}')~
w_{\alpha}^{(j)}({\bf r}-{\bf r}') \;,
\end{equation}
with geometrical weight functions $w^{(j)}_{\alpha}$ of species $j=1,\ldots,N$.
In Rosenfeld's approach there are six different weighted densities, four are 
scalar and two are vector-like. For a general hard body $j$ one has 
\cite{Rosenfeld94,Rosenfeld95} 
\begin{eqnarray}
w^{(j)}_{3}({\bf r})&=&\Theta(|{\bf r}-{\bf R}^{(j)}|)\;,\nonumber\\
w_{2}^{(j)}({\bf r})&=&\delta (|{\bf r}-{\bf R}^{(j)}|)\;,\nonumber\\
{\bf w}_{V2}^{(j)}({\bf r})&=&\hat{{\bf n}}^{(j)}\: \delta(|{\bf r}-{\bf R}^{(j)}|)\;,\nonumber\\
w_{1}^{(j)}({\bf r})&=&\frac{H^{(j)}}{4\pi}w_{2}^{(j)}({\bf r})\;,\nonumber\\
w_{0}^{(j)}({\bf r})&=&\frac{K^{(j)}}{4\pi}w_{2}^{(j)}({\bf r})\;,\nonumber\\
{\bf w}_{V1}^{(j)}({\bf r})&=&\frac{H^{(j)}}{4\pi}{\bf w}_{V2}^{(j)}({\bf r})\;,\label{eqn:22}
\end{eqnarray}
where ${\bf R}^{(j)}={\bf R}^{(j)}(\theta,\varphi)$ is the radius vector to 
the surface and $\hat{{\bf n}}^{(j)}$ is its outward unit vector normal to the
surface at point ${\bf R}^{(j)}$ of particle $j$. $H^{(j)}$ and $K^{(j)}$ 
are the integrated mean curvature and integrated Gaussian curvature of the 
particle $j$, respectively \cite{Rosenfeld94,Rosenfeld95}. 

In the bulk limit for which the density profiles $\rho_{j}({\bf r})$ are 
constant, as considered in this section, the vector weighted densities vanish,
and  Eq.~(\ref{eqn:15}) takes the form
\begin{equation}\label{eqn:23}
-c^{(1)}_{b}= 
\frac{\partial \Phi}{\partial n_{3}}~\zeta_3+
\frac{\partial \Phi}{\partial n_{2}}~\zeta_2+
\frac{\partial \Phi}{\partial n_{1}}~\zeta_1+
\frac{\partial \Phi}{\partial n_{0}}~\zeta_0 \;,
\end{equation} 
with characteristic functions $\zeta_{i}$, $i=0,\ldots ,3$ of the shape of the
particle of species $b$.

Assuming the general functional of the given by 
Eqs.~(\ref{eqn:20})-(\ref{eqn:22}),
one has for a general convex hard particle \cite{Rosenfeld94,Rosenfeld95} 
\begin{eqnarray}
\zeta_{3} & = & V^{(b)},\label{eqn:29}\\
\zeta_{2} & = & A^{(b)}, \label{eqn:30}\\
\zeta_{1} & = & A^{(b)} \frac{H^{(b)}}{4 \pi}, \label{eqn:31}
\end{eqnarray}
and
\begin{equation}
\zeta_{0}  =  A^{(b)} \frac{K^{(b)}}{4 \pi} \label{eqn:32},
\end{equation}
where the shape of the particle is specified by its volume $V^{(b)}$, its
surface area $A^{(b)}$, and integrated curvatures $H^{(b)}$ and $K^{(b)}$.

Within FMT there are several expressions for the excess free energy density 
$\Phi$. For the given problem we have chosen the original Rosenfeld form 
\cite{Rosenfeld89}
\begin{equation}\label{eqn:24}
\Phi(\{n_{\alpha}\})=-n_{0} \ln (1-n_{3})+\frac{n_{1}n_{2}-{\bf n}_{1}\cdot 
{\bf n}_{2}}{1-n_{3}} +\frac{n_{2}^{3}-3n_{2}{\bf n}_{2}\cdot{\bf n}_{2}}
{24\pi (1-n_{3})^{2}} \;.
\end{equation} 
The partial derivatives of the excess free energy density with respect to the 
weighted densities in the dilute limit of species $b$ depend only on the 
packing fraction $\eta$ and the radius $R$ of the hard-sphere fluid:
\begin{eqnarray}
\frac{\partial \Phi}{\partial n_{3}} & = & \frac{3 \eta}{4 \pi R^3}
\frac{1+\eta+\eta^2}{(1-\eta)^3}\equiv\beta p_{PY} \label{eqn:25}  \;,\\
\frac{\partial \Phi}{\partial n_{2}} & = & \frac{3 \eta (2+\eta)}
{8 \pi R^2 (1-\eta)^2}\equiv\beta \gamma_{HW}^{SPT} \label{eqn:26}\;,\\
\frac{\partial \Phi}{\partial n_{1}} & = & \frac{3 \eta}{R (1-\eta)}
\equiv\phi_1 \label{eqn:27}\;,
\end{eqnarray}
and
\begin{equation}
\frac{\partial \Phi}{\partial n_{0}}  =  -\ln (1-\eta)\equiv\phi_0 \;.
\label{eqn:28}
\end{equation}
In the above equations $p_{PY}$  and $\gamma^{SPT}_{HW}$ are the Percus-Yevick
compressibility equation of state and the scaled-particle theory (SPT) 
expression for the surface tension of a hard-sphere fluid at a planar hard 
wall \cite{Reiss60}, respectively.

With the characteristic functions in Eqs.~(\ref{eqn:29})-(\ref{eqn:32}) and 
the partial derivatives given by Eqs.~(\ref{eqn:25})-(\ref{eqn:28}) one
finds within this approach for the grand potential of a hard-sphere fluid
surrounding any convex hard particle {\it b}
\begin{equation} \label{eqn:33}
\beta\Omega = -\beta p_{PY} (V_{tot}-V^{(b)}) + 
A^{(b)} \left(\beta\gamma_{HW}^{SPT}+\frac{H^{(b)}}{4 \pi}~
\phi_1 + \frac{K^{(b)}}{4 \pi}~ \phi_0 \right) \;.
\end{equation} 

By choosing the surface of the convex particle as the dividing interface, the 
bulk contribution to the grand potential is simply 
$\Omega_0=-p^{PY} (V_{tot}-V^{(b)})$ and the surface tension follows directly 
as 
\begin{equation} \label{eqn:34}
\beta\gamma(H^{(b)},K^{(b)}) =\beta\gamma_{HW}^{SPT}+\frac{H^{(b)}}{4 \pi}~\phi_1 + 
\frac{K^{(b)}}{4 \pi}~ \phi_0 \;.
\end{equation}
The particular choice of the dividing interface underlying Eq.~(\ref{eqn:34}) 
is motivated by the simplicity of the resulting expression for the surface 
tension. Other choices of the dividing interface would lead to significantly 
more complex expressions and the elegant, simple form of Eq.~(\ref{eqn:34}) 
would be lost. Obviously within this bulk approach it is impossible to pick up
a contribution to the surface tension that is proportional to the logarithm of
the radius of curvature. 

The surface tension in Eq.~(\ref{eqn:34}) is always positive for a non-zero 
packing fraction $\eta$ and its change for small curvature due to an increase 
in curvature is described by the Tolman length $\delta_T$ which within this
approach is defined by
\begin{equation}\label{eqn:35}
\gamma(H^{(b)},K^{(b)})= \gamma_{HW}^{SPT} \left(1-2 H^{(b)} \delta_T
+{\cal O}(K^{(b)})\right) \; ,
\end{equation}
so that it is independent of the shape of the convex particle with
\begin{equation} \label{eqn:tolman}
\delta_T = -\frac{\phi_1}{8\pi\beta\gamma_{HW}^{SPT}} = {\frac{\eta - 1}{\eta + 2}} R \;.
\end{equation}

Thus within this approach we find that the generalized Tolman length for a general
hard convex cavity is negative, provided the surface of this hard cavity is chosen
as the dividing surface. The grand potential (Eq.~(\ref{eqn:33})) and the surface tension
(Eq.~(\ref{eqn:34})) have been derived for a general convex hard wall and
are related to the Helfrich Hamiltonian \cite{Helfrich73}.
In order to be able to calculate the contact density of a hard-sphere fluid
on the basis of the sum rule in Eq.~(\ref{eqn:7}) one has to evaluate
the expression in Eq.~(\ref{eqn:34}) for special cases. For a big spherical particle the mean 
curvature $H_s^{(b)}=1/R_s^{(b)}$ and the Gaussian curvature 
$K_s^{(b)}=1/(R_s^{(b)})^2$. By combining Eqs.~(\ref{eqn:34}) and 
(\ref{eqn:10}) within the present bulk theory  one finds for the contact value
of the density profile of a hard-sphere fluid at a big hard sphere:
\begin{equation}\label{eqn:conts}
\rho(R_s^+)= \beta p_{PY} - \frac{9 \eta^2}{4 \pi (1-\eta)^3R^{3}}
\left\{(1+\eta)\frac{R}{R_{s}}-\eta\frac{R^{2}}{R_{s}^{2}}\right\}\;.
\end{equation}

For finite curvatures the contact density is smaller than in the planar limit,
which in the present approach is given by the Percus-Yevick compressibility 
pressure $\beta p_{PY}$. The difference between the contact density at finite 
curvature and its planar wall limit is a quadratic function in $1/R_s$.

For a cylindrical wall we have to proceed more carefully. Since 
Eq.~(\ref{eqn:34}) was derived assuming a finite convex body we take the limit
$L \to \infty$ only after the cylinder is inserted into the bulk system. In 
this limit the mean curvature $H_c^{(b)}=1/(2 R_c^{(b)})$ and the Gaussian 
curvature $K_c^{(b)}=0$. The contact density of a hard-sphere fluid close to a
cylindrical wall follows from Eq.~(\ref{eqn:34}) and the sum rule in 
Eq.~(\ref{eqn:11}) as
\begin{equation}\label{eqn:contc}
\rho(R_c^+)=\beta p_{PY}-\frac{9 \eta^2}{4 \pi (1-\eta)^3  R^3}
\left\{\frac{(1+\eta)}{2}\frac{R}{R_{c}}
\right\}\;,
\end{equation}
which is linear in $1/R_c$. For large radii of curvature $R_i\to \infty$ the 
contact densities at a big sphere $\rho(R_s^+)$ and at a cylinder $\rho(R_c^+)$
are related by $2\rho(R_c^+)-\rho(R_s^+) = \beta p_{PY}+{\cal O}(\frac{1}{R_{s}^{2}}) $.

\section{Numerical results} \label{sec:DFT}

As mentioned in the Introduction an alternative approach for calculating the 
surface tension and the contact density of a hard-sphere fluid at a spherical 
or cylindrical wall consists of minimizing the density functional, 
Eq.~(\ref{eqn:5}), numerically. Within this approach the curved wall enters 
the problem via the external potential specified in Eq.~(\ref{eqn:1}). The 
solution of the corresponding Euler-Lagrange equation yields the inhomogeneous
equilibrium density profile $\rho(r)$. From $\rho(r)$ both the contact 
densities $\rho(R_i^+)=\rho(r=R_i^+)$ and the surface tensions 
$\gamma_i(R_i^{(b)})$, Eq.~(\ref{eqn:3}), follow directly. In order to verify 
that the numerical DFT results for the contact density and the surface tension
(from minimization) are consistent and to probe the numerical accuracy of our
DFT calculation, we confirmed numerically that the sum rule in 
Eq.~(\ref{eqn:7}) is satisfied very accurately (within the range of four to 
six significant digits depending on the packing fraction).

Although the minimization of the density functional and the bulk route 
discussed in Sec.~\ref{sec:bulk} are in principal equivalent approaches, they 
differ in practice because the actual density functional for the hard-sphere 
fluid is an approximation. Within an approximate treatment it is expected that
the numerical solution of the Euler-Lagrange equation, which is based on a 
variational principle, is more accurate than the aforementioned bulk approach.
Nevertheless, for the hard-sphere fluid which, apart from freezing, 
exhibits only a single phase both routes are expected to be 
in qualitative or even semi-quantitative agreement with each other.

First we present in Fig.~\ref{fig:1} our numerical results for the 
surface tension at a spherical wall as obtained by full minimization of the
Rosenfeld functional for  values of $\eta$ in the range between $\eta=0.1$ and
$\eta=0.42$ and with radii of curvature $R_s^{(b)}$ in the range $R_s^{(b)}=R$
and $R_s^{(b)}=10000 R$. The DFT results are denoted by symbols and the 
predictions of the bulk theory by full lines. For small values of $\eta$ we 
find both results to be in excellent, quantitative agreement for all 
curvatures, see Fig.~\ref{fig:1}(a). Close to the planar wall limit the level 
of quantitative agreement reduces significantly for larger packing fractions, 
while it remains good for large curvatures. This deviation between the two 
different routes is expected because in the limit $R^{(b)}_{s}\to\infty$ 
the bulk route reduces to the SPT prediction of the surface tension at a 
planar wall, which is known to overestimate the surface tension of a 
hard-sphere fluid at a planar hard wall \cite{Roth01b}. Since SPT obtains the 
planar wall limit from extrapolating the behavior at small cavity sizes 
\cite{Reiss60} this deviation is no surprise. Recently it was suggested that 
SPT can be re-optimized by taking the planar wall limit into account more 
accurately \cite{Henderson02}.

Even for packing fractions at which the level of quantitative agreement is
moderate  (see Fig.~\ref{fig:1}(b)), there is a good qualitative agreement between the
DFT results and our bulk approach. This implies that the 
curvature dependence of our numerical data of $\gamma_{s}$ can be best 
approximated by a polynomial quadratic in $1/R_s^{(b)}$ with no evidence for 
a logarithmic term.

The DFT results for the surface tension of a hard-sphere fluid in contact with
a hard cylindrical wall are shown in Fig.~\ref{fig:2}. Again the quantitative
agreement between the DFT results (symbols) and predictions of the bulk theory
(full lines) are excellent for low packing fractions and fair at higher values
of $\eta$. The curvature dependence of the DFT data is captured very well by a
function linear in $1/R_c^{(b)}$ as predicted by the bulk route.

A remarkable feature of Eq.~(\ref{eqn:34}) is that the ratio of the 
coefficients corresponding to the term of the leading order in $1/R_{i}$ for 
spherical and cylindrical symmetry is exactly 2 (since 
$H_{s}^{(b)}=2H_{c}^{(b)}$). This behavior can also be found in the numerical 
results  of DFT calculations, i.e., based on the inhomogeneous density profiles
$\rho (r)$. In Table~\ref{tab:1} we show values of least-square fits assuming
that $\beta\gamma_{s}(R_{s}^{(b)})=a_{s}^{(0)}+\frac{a_{s}^{(1)}}{R_{s}^{(b)}}+
\frac{a_{s}^{(2)}}{(R_{s}^{(b)})^{2}}$
and $\gamma_{c}(R_{c}^{(b)})=a_{c}^{(0)}+\frac{a_{c}^{(1)}}{R_{c}^{(b)}}$,
i.e., the simplest formulae to fit our DFT data. To a good approximation the 
ratio $a_{s}^{(1)}/a_{c}^{(1)}$ equals 2 for all densities. This ratio agrees 
with the predictions of the bulk theory; $a_{s}^{(0)}=a_{c}^{(0)}$ is the 
planar wall surface tension. The Tolman length (compare Eq.~(\ref{eqn:35}))
is given by $\delta_T^{(s)}=-a_s^{(1)}/(2 a_s^{(0)})$ and 
$\delta_T^{(c)}=-a_c^{(1)}/a_c^{(0)}$. In Fig.~\ref{fig:tolman} we plot
$\delta_T^{(s)}$ (diamonds) and $\delta_T^{(c)}$ (circles) together with the
prediction of the bulk theory Eq.~(\ref{eqn:tolman}) (full line) as a function
of the packing fraction $\eta$. 

Thus neither the bulk theory nor the numerical DFT results provide any hints 
for the occurrence of logarithmic terms in the expression of the surface 
tension of hard-sphere fluids in contact with curved hard surfaces.

For the contact density one knows a priori that for all values of $\eta$ both 
the DFT route and the bulk approach yield the same contact value 
$\rho (R_{i}^{+}\to\infty)=\beta p$ of the density profile of a hard-sphere 
fluid in the limit of the planar wall. For finite curvature we find that the 
contact densities obtained numerically from DFT (symbols) and analytically from
our bulk theory for $\gamma (R_{s}^{(b)})$ (full lines) are in very good 
quantitative agreement for all values of $\eta$, as shown in Fig.~\ref{fig:3} 
for spherical walls and in Fig.~\ref{fig:4} for cylindrical walls. Only for 
very large curvatures the actual DFT contact densities are slightly 
overestimated by the bulk theory.

For the spherical wall in Fig.~\ref{fig:3} we also plot the contact density
obtained from an empirical fit to simulation data (dotted lines) by Degreve et
al. \cite{Degreve94}. In terms of $R_{s}^{-1}$ this fit interpolates linearly
between the planar wall limit within the Carnahan-Starling theory and the 
contact value of the pair correlation function $g(r)$ corresponding to
$R/R_s=0.5$. At high packing 
fractions this fit deviates from our results in the planar wall limit due to 
the difference between the Carnahan-Starling equation of state and the 
Percus-Yevick compressibility pressure which underlies both our approaches. At
very high packing fractions there are additional deviations because the simple
linear fit does not capture the actual higher order terms in $R_{s}^{-1}$. 
Nonetheless, the overall agreement between our results and this fit is good.

\section{Conclusions and Summary} \label{sec:conclusion}

\begin{figure}
\vspace{0.5cm}
\centering\epsfig{file=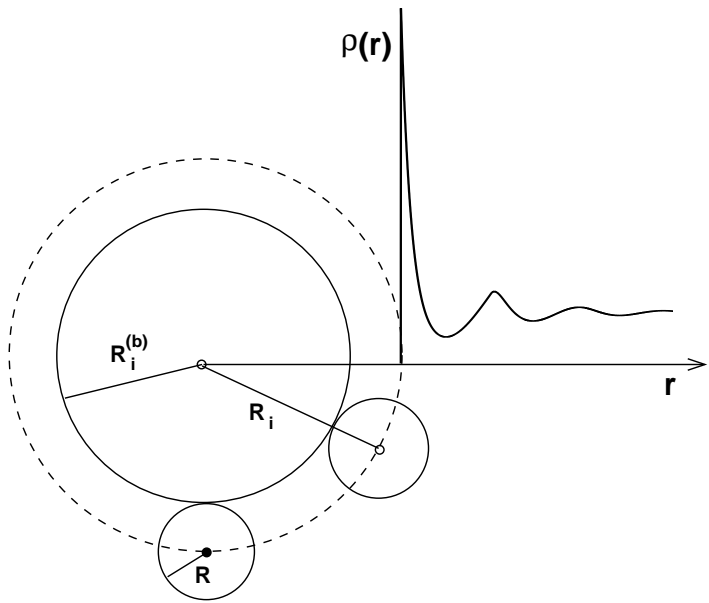,width=12cm}
\vspace{0.5cm}
\caption{\label{fig:geometry} Geometry of a hard-sphere fluid in contact
with a hard spherical ($i=s$) or cylindrical ($i=c$) wall with radius 
$R_i^{(b)}=R_i-R$ and volume $V^{(b)}$. While the density profile $\rho(r)$
attains its contact value at $r=R_i$, we choose the wall, i.e., the surface 
with radius $R_i^{(b)}$ as dividing surface for the calculation of the surface
tension.}
\end{figure}

\begin{figure}
\vspace{0.5cm}
\centering\epsfig{file=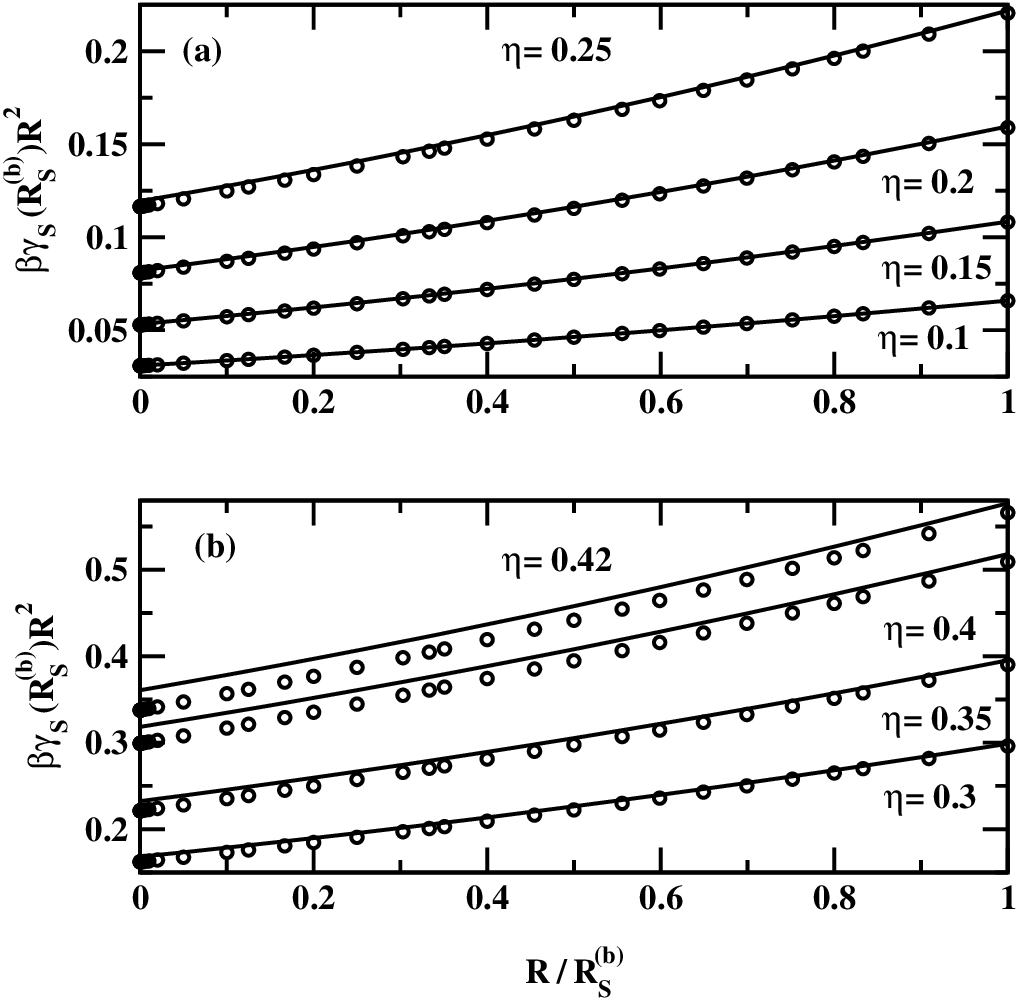,width=12cm}
\vspace{0.5cm}
\caption{\label{fig:1} The surface tension of a fluid of hard spheres with 
radius $R$ near a hard 
spherical wall of radius $R_s^{(b)}$. Symbols denote results obtained from 
direct minimization of the Rosenfeld functional, whereas the full lines are 
predictions of the bulk theory (see Eq.~(\ref{eqn:34})). Small packing fractions
($\eta =0.1,\ldots, 0.25$) are shown in (a) and large packing fractions 
($\eta=0.3,\ldots, 0.42$) in (b). The agreement between both routes is 
excellent for low packing fractions. Towards high values of $\eta$ the level 
of quantitative agreement decreases, while the bulk theory still reproduces 
qualitatively the DFT results.}
\end{figure}

\begin{figure}
\centering\epsfig{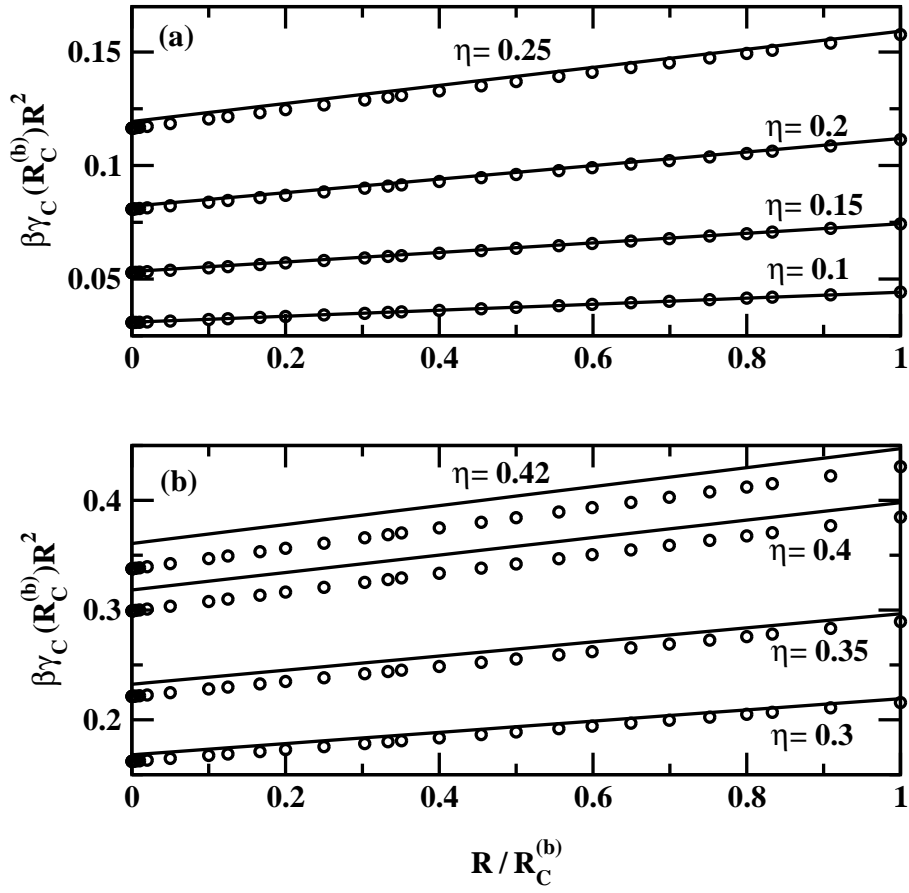}
\vspace{0.5cm}
\caption{\label{fig:2} 
The surface tension of a fluid of hard spheres with radius $R$ near a hard cylindrical wall of 
radius $R_c^{(b)}$. Symbols denote results obtained from direct minimization 
of the Rosenfeld functional, whereas the full lines are predictions of the 
bulk theory (see Eq.~(\ref{eqn:34})). Small packing fractions 
($\eta =0.1,\ldots, 0.25$) are shown in (a) and large packing fractions 
($\eta=0.3,\ldots, 0.42$) in (b). The agreement between both routes is 
excellent for low packing fractions. Towards high values of $\eta$ the level 
of quantitative agreement decreases, while the bulk theory still reproduces 
qualitatively the DFT results.}
\end{figure}

\begin{figure}
\vspace{0.5cm}
\centering\epsfig{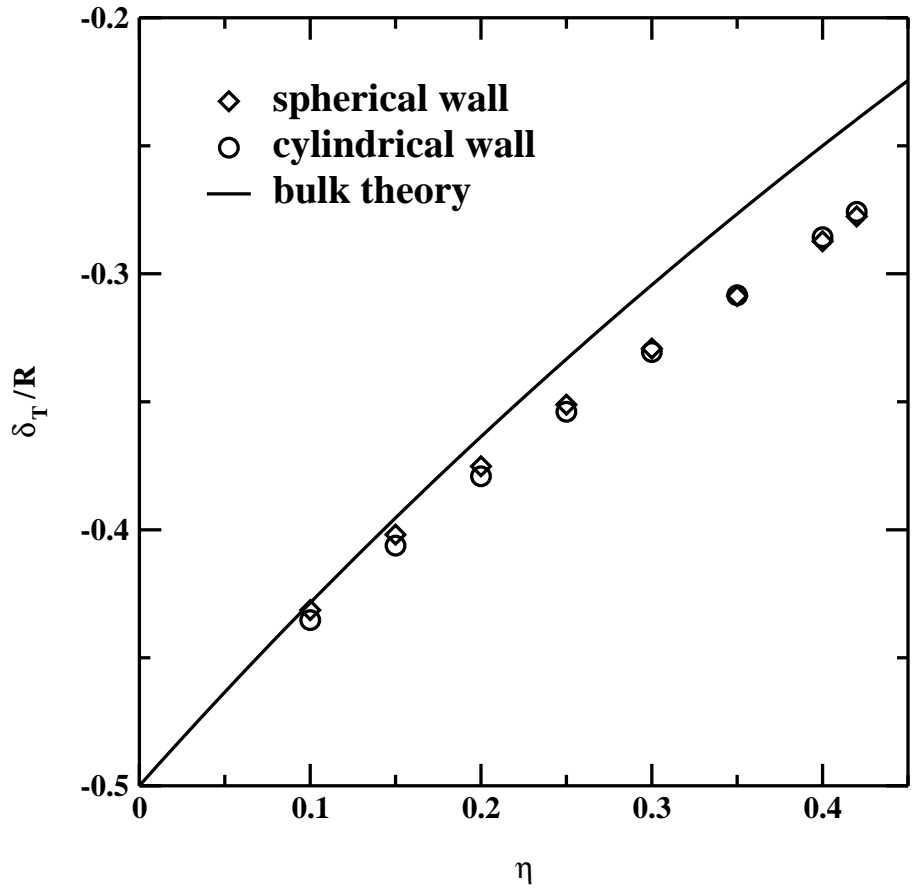}
\vspace{0.5cm}
\caption{\label{fig:tolman} The Tolman length $\delta_T$ of a fluid
of hard spheres with radius $R$
at curved hard walls as a function of the packing fraction $\eta$. Diamonds
and circles denote results for spherical and cylindrical walls, respectively,
while the full line correspond to the prediction of the bulk theory 
(Eq.~(\ref{eqn:tolman})). 
}
\end{figure}

\begin{figure}
\centering\epsfig{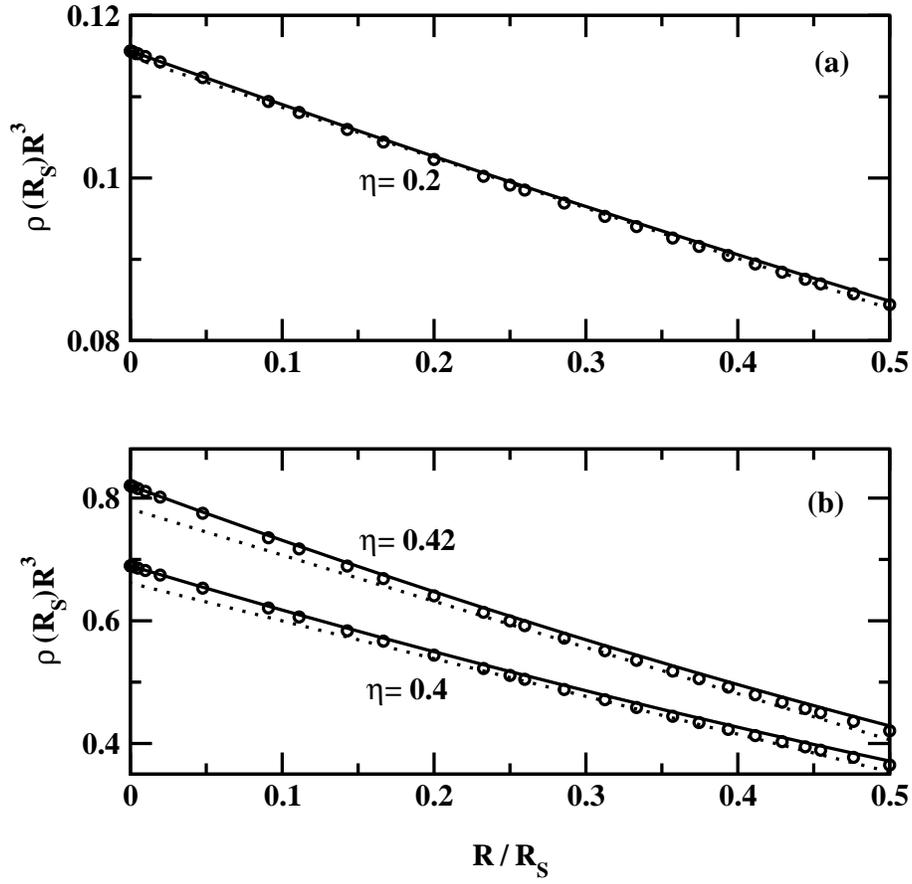}
\vspace{0.5cm}
\caption{\label{fig:3} The contact values of the density profile of a 
fluid of hard spheres with radius $R$ at a hard spherical wall with 
radius $R_s^{(b)}=R_s-R$ as obtained
from direct minimization of the Rosenfeld functional (symbols), from the bulk
theory (Eq.~(\ref{eqn:conts})) (solid lines), and from a semi-empirical 
parameterization \protect\cite{Degreve94} (dotted lines). The results for the 
packing fraction $\eta=0.2$ are given in (a) and for $\eta=0.4$ and 
$\eta=0.42$ in (b). Even for high packing fractions the agreement between the 
DFT results and the predictions of the bulk theory is very good.}
\end{figure}

\begin{figure}
\centering\epsfig{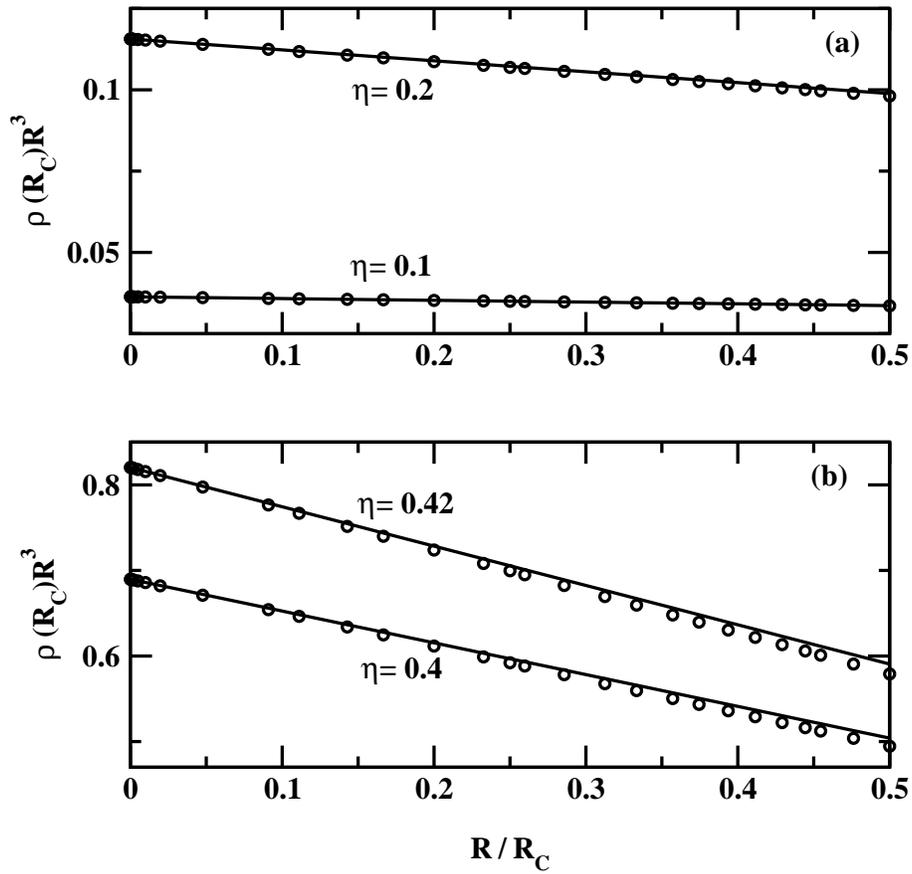}
\vspace{0.5cm}
\caption{\label{fig:4} 
The contact values of the density profile of a fluid of hard spheres 
with radius $R$ at a hard 
cylindrical wall with radius $R_c^{(b)}=R_c-R$ as obtained from direct minimization 
of the Rosenfeld functional (symbols) and from the bulk theory 
(Eq.~(\ref{eqn:contc})) (solid lines). The results for the packing fractions 
$\eta=0.1$ and $\eta=0.2$ are given in (a) and for $\eta=0.4$ and 
$\eta=0.42$ in (b). Even for
high packing fractions the agreement between the DFT results and the 
predictions of the bulk theory is very good.}
\end{figure}

Our analysis of the local structure and the thermodynamics of the hard-sphere 
fluid near hard curved substrates has lead to the following main results:
\begin{enumerate}
\item We have studied the curvature dependence of the surface tension of a 
hard-sphere fluid at hard spherical and cylindrical walls 
(Fig.~\ref{fig:geometry}) obtained by minimizing the Rosenfeld fundamental 
measure density functional. We have found no indications of logarithmic 
singularities in the expansion of the surface tension in terms of curvatures of
spherical or cylindrical walls.  The surface tension appears to be an analytic
function of curvature.

\item Based on the Rosenfeld fundamental measure theory we have derived an 
analytical expression for the surface tension of a hard-sphere fluid close to 
hard, arbitrarily curved, convex walls in terms of their integrated mean and 
Gaussian curvatures, $H^{(b)}$ and $K^{(b)}$ (see Eq.~(\ref{eqn:34})). This 
approach also does not render logarithmic singularities of the surface tension
as a function of the radii of curvature. There is good agreement between the 
results of the density functional theory and the bulk theory which treats the 
curved wall as the surface of a particle of a second component in infinite 
dilution (Figs.~\ref{fig:1} and \ref{fig:2}). The Tolman length as a function
of the packing fraction of the hard-sphere fluid is shown in Fig.~\ref{fig:tolman}.

\item Based on the sum rule for curved substrates we have obtained expressions 
for the contact values of the density profile of a hard-sphere fluid close to 
spherical (Eq.~(\ref{eqn:conts})) and cylindrical (Eq.(\ref{eqn:contc})) walls.
A comparison with the density functional results indicates, that the 
analytical expressions of the contact values are very reliable; only at rather
high packing fractions they slightly overestimate the density functional 
values (Figs.~\ref{fig:3} and \ref{fig:4}) for large curvatures.

\item The ratio of the leading order terms in the expansion of the surface 
tension as a function of curvature, obtained from full minimization of the 
fundamental measure functional, for cylindrical and spherical symmetry equals 
2 (see Table~\ref{tab:1}). This is in agreement with the aforementioned bulk 
theory (Eq.~(\ref{eqn:34})). This also agrees with the general feature of the 
Helfrich theory \cite{Helfrich73}, in which for a sphere the contribution to 
the surface tension linear in the curvature is twice as large as  for a 
cylinder.
\end{enumerate}

In Ref.~\cite{Rowlinson94} the arguments in favor of and against the existence
of logarithmic singularities in the curvature dependence of the surface 
tension of fluids have been reviewed. Our present results indicate that for a 
hard-sphere fluid near hard curved substrates such logarithmic singularities
either do not exist or their influence is negligible.

It would be of interest to investigate the issue of analyticity of the surface
tension via a diagrammatic expansion of the contact value of the density 
profile \cite{Nijboer52,Stecki78}. Within this approach our preliminary 
calculations show, that in the lowest order in density the diagrammatic 
expansion of the contact value of the density profile indeed leads 
to a surface tension analytical in curvature.
However, the extension of this analysis to higher orders in densities is 
technically involved and is postponed to future work. 

Thus our findings indicate that non-analyticities of the curvature dependence
of surface tensions arise only via dispersion forces acting on the fluid
particles \cite{Bieker98} or via the onset of drying or wetting transitions
on curved substrates \cite{Evans03}

\acknowledgments
It is a pleasure to thank Bob Evans for stimulating discussions.

\newpage
\begin{table}
\caption{\label{tab:1} Values of $a_{s}^{(0)}$, $a_{s}^{(1)}$, $a_{s}^{(2)}$ 
and $a_{c}^{(1)}$ as obtained from least square fit to DFT data of the surface
tension of a hard-sphere fluid in contact with a spherical and cylindrical
hard wall for various values of the packing fraction $\eta$. Note that 
Eq.~(\ref{eqn:34}) predicts $\frac{a_{s}^{(1)}}{a_{c}^{(1)}}=2$ for all $\eta$.
}
\begin{tabular}{cccccc}
$\eta$& $a_{s}^{(0)}R^{2}=a_{c}^{(0)}R^{2}$&$a_{s}^{(1)}R$&$a_{s}^{(2)}$&$a_{c}^{(1)}R$&$a_{s}^{(1)}/a_{c}^{(1)}$\\ \tableline
0.10    &   0.03083  &    0.02660   &   0.00837  &  0.01342 & 1.98 \\
0.15    &   0.05281  &    0.04245   &   0.01286  &  0.02145 & 1.98 \\
0.20    &   0.08078  &    0.06061   &   0.01754  &  0.03062 & 1.98 \\
0.25    &   0.11643  &    0.08175   &   0.02235  &  0.04121 & 1.98 \\
0.30    &   0.16213  &    0.10674   &   0.02725  &  0.05360 & 1.99 \\
0.35    &   0.22127  &    0.13660   &   0.03229  &  0.06823 & 2.00 \\
0.40    &   0.29924  &    0.17195   &   0.03800  &  0.08548 & 2.01 \\
0.42    &   0.33761  &    0.18747   &   0.04084  &  0.09312 & 2.01 \\
\end{tabular}
\end{table}

\end{document}